\documentclass[aps,twocolumn]{revtex4}

\usepackage{amssymb}
\usepackage{epsfig}

\begin{document}

\title{Threshold of coexistence and critical behavior 
of a predator-prey cellular automaton}
\author{Everaldo Arashiro
and T\^{a}nia Tom\'{e}}
\affiliation{Instituto de F\'{\i}sica \\
Universidade de S\~{a}o Paulo\\
Caixa postal 66318\\
05315-970 S\~{a}o Paulo - SP, Brazil}
\date{\today}

\begin{abstract}
We study a probabilistic cellular automaton to describe two population
biology problems: the threshold of species coexistence in a predator-prey
system and the spreading of an epidemic in a population. By carrying out
time-dependent simulations we obtain  the dynamic critical
exponents and the phase boundaries (thresholds) related to
the transition between an active
state, where prey and predators present a stable coexistence, and a prey
absorbing state. The estimates for the critical exponents show that the
transition belongs to the directed percolation universality class. In the
limit where the cellular automaton maps into a model for the spreading of an
epidemic with immunization we observe a crossover from directed percolation
class to the dynamic percolation class. Patterns of growing clusters related
to species coexistence and spreading of epidemic are shown and discussed.

\end{abstract}

\maketitle

\section{Introduction}

In 1958, Huffaker \cite{huffaker}, in a pioneering experiment, has been able
to maintain in the laboratory a population of prey and predators
coexisting and presenting self-sustained coupled time oscillations.
He has verified
that persistence of species was only possible in a large and heterogeneous
space. Since then different models have been proposed to explain the 
r\^{o}le of space in determining the species coexistence 
\cite{durrett1,tainaka,sat,bocara1,provata,
droz,lipowsky,ilka,aguiar,kelly,szabo,tauber,tpb}. A common feature of these
models is that they are based either on interacting particle systems 
\cite{liggett,durrett0}, also called stochastic lattice
models \cite{sat}, or on probabilistic cellular automata.
These descriptions are
appropriate and relevant when considering predator-prey systems
which are under conditions of very low species population densities
and/or when their habitat is spatially heterogeneous \cite{durrett1}.

In the present article we study a stochastic model that takes
into account the spatial structure explicitly to describe
two population biology issues: the
threshold of species coexistence in a predator-prey system
and the spreading of an infectious disease in a population.
The individuals of each population
are treated as discrete and they are supposed to occupy the sites of a
two-dimensional lattice. The interactions between individuals are included
in the local stochastic rules which take into account the Lotka-Volterra
mechanisms of interaction \cite{lotka,volterra}.
The dynamics associated to
the system is a Markovian discrete time process
defined by a probabilistic cellular automaton, 
having three states per site,
which we call predator-prey cellular
automaton.
The connection between pattern formation
and species coexistence in this model was recently analyzed
\cite{tpb}.
Results from steady-state simulations
have shown that, depending on the set of the parameters of the
model, the following steady states can be attained \cite{tpb}: 
a prey absorbing state, where
predators have been extinct and the lattice is full of prey,
and an active state where both species coexist with constant
time densities. For finite
systems, it was also detected an active state
where both population densities oscillate in time
with the same characteristic frequency \cite{kelly}.

Our aim in this work is to determine with precision the threshold of stable
coexistence of species in the predator-prey cellular automaton.
To this end we perform time-dependent simulations 
\cite{grassberger,grassberger1,grassberger2,grassberger89,
grassberger3,dickman1,dicktt,jensen,hinrichsen,marro,mun,silva,ed}
which allow us to obtain the
phase boundaries together with the critical exponents.
The scaling analysis of the time-dependent
simulations yields a set of dynamic critical exponents
and thus the possibility of
classification of the models with absorbing states
in universality classes.
Here, we have found that the automaton exhibits a line
of continuous phase transition which belongs to the
universality class of directed percolation, for all sets of
parameters such that the prey birth
probability is different from zero. 
When this probability vanishes there occurs a
crossover to the universality class of dynamic isotropic
percolation.
In this limit the predator-prey cellular automaton is
mapped into a model for the spreading of
an epidemic with immunization, a
general epidemic process 
\cite{grassberger1,grassberger2,bocara,hastings,hinrichsen1}.

This article is organized as follows. In Sec. II we describe the
predator-prey cellular automaton and the time-dependent simulations.
Critical properties and pictures of growing clusters generated by the
simulations are shown in the same section. In Sec. III we analyze
the critical behavior of a model for an epidemic spreading. We briefly
discuss and summarize our results in Sec. IV.

\section{Predator-prey probabilistic cellular automaton}

\subsection{The model}

We assume that each individual of each species population
can reside on the sites of a regular square lattice
which represents their habitat.
A site in
the lattice can be in one of three states: occupied by a prey
individual ($X$); occupied by a predator individual ($Y$) or empty ($Z$). 
The predator-prey
probabilistic cellular automaton comprehends the following processes: 
\begin{equation}
Z+X\rightarrow 2X,   
\label{1}
\end{equation}
\begin{equation}
X+Y\rightarrow 2Y,  
\label{2}
\end{equation}
and 
\begin{equation}
Y\rightarrow Z.  
\label{3}
\end{equation}
The transitions between states obey probabilistic rules which depend on the
state of the given site and on the states of its four nearest neighbors at
the north, east, south and west (which defines the neighborhood). 
The update is synchronous and the rules are:

(i) A prey individual can be born in an empty site
with a probability $a/4$ times the number of
sites occupied by prey in the neighborhood,
process (\ref{1}).

(ii) A predator individual can be born in a site occupied by a prey if there
are predators in its neighborhood. Prey disappear instantaneously and
give place to a new predator. The probability of this process is $b/4$ times
the number of sites occupied by predators in the neighborhood,
process (\ref{2}).

(iii) The death of predators is spontaneous: 
a site occupied by a predator
can be evacuated with a probability $c$. This process reintegrates to the
system the resources for prey proliferation, process (\ref{3}).

\begin{figure}
\centering\epsfig{file=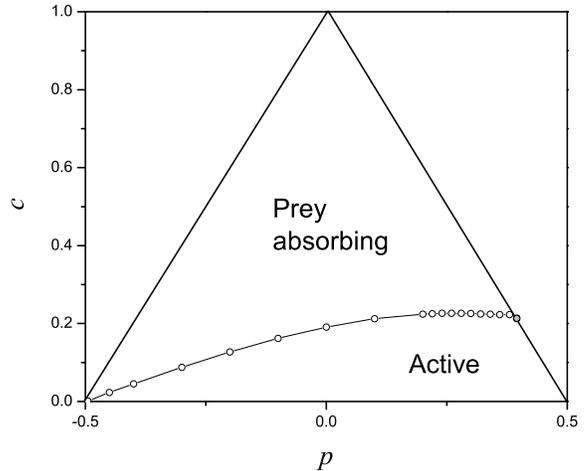,width=9.0cm}
\caption{Phase diagram in the plane $p-c$, showing regions corresponding to
the prey absorbing phase and the active phase. Phases are separated by a
transition line obtained
from time-dependent simulations (see Table I).}
\label{diag}
\end{figure}

The model has three parameters $a$, $b$, and $c$, with $0\leq a\leq 1$, 
$0\leq b\leq 1$ and $0\leq c\leq 1$. However, we assume that $a+b+c=1$, thus
just two parameters are independent. We consider the parametrization 
\cite{sat,tpb}: $a=(1-c)/2-p$ and $b=(1-c)/2+p$. 
The parameter $p$ is such that $-1/2\leq p\leq 1/2$ and parameter $c$ 
denotes the death probability of
predators, $0\leq c\leq 1$.
This parametrization allows us to analyze the
model in a triangular phase diagram $p-c$, as shown in Fig. \ref{diag}.

The system evolves in time and can eventually reach stationary states
characterized by constant time densities of prey, predators, and empty
sites.
These can be either active states where
prey and predators have
stable coexistence or absorbing states.
We note that finite systems also exhibit
local oscillatory behavior of the densities of prey and predators 
\cite{kelly}. 

A transition line $c_{1}(p)$ crosses the
entire $p-c$ diagram represented in Fig. \ref{diag}.
It starts from the left corner of the triangle and ends on
the opposite side. The densities of prey, predators and empty sites change
continuously at $c_{1}(p)$. For $a\neq 0$, the transition is from the active
state to the prey absorbing state and the transition line was estimated
firstly in reference \cite{tpb}. In the limiting case 
corresponding to $a=0$, the point where $c_{1}(p)$ 
meets the right side of triangle of Fig. \ref{diag},
the transition is from one of the infinitely
many absorbing states to the prey absorbing state. In order to determine the
critical line $c_{1}(p)$, with precision, and the dynamic critical
exponents, we perform time-dependent simulations.

\subsection{Critical exponents and thresholds}

To carry out a time-dependent simulation analysis for the predator-prey
cellular automaton we follow the time evolution of states very close
to the prey absorbing state. 
We depart from an initial condition ($t=0$) with a single
predator at the origin of a lattice covered by prey. Once the system is
placed in the initial condition we apply the local rules (i), (ii), and
(iii), described in Sec. II A, and let it evolves in time. We have
considered $N_{S}=10000$ samples (independent runs) all starting with this
same condition. For each fixed value of $p$ we vary $c$ near its transition
value. The simulations were performed in systems sufficiently large so that
predators do not reach the borders of the lattice.

We have investigated the behavior of the following quantities: the survival
probability of predators $P$, that is, the probability that predators have
not been extinct until time $t$; the mean-number of predators $\langle
N\rangle $ at time $t$, whose average is calculated over all the samples,
including those where predators have been extinct before time $t$; and
the average mean-square distance of spreading of predators from the origin,
$\langle R^{2}\rangle $. This average is calculated by
taking into account only the samples that survived until time $t$.

According to the scaling laws for
time-dependent simulations \cite{grassberger},
for large values of $t$, we expect, at the critical point, the following
power laws, 
\begin{equation}
\langle N\rangle ~\sim t^{\eta },  \label{sca1}
\end{equation}
\begin{equation}
P\sim t^{-\delta },  \label{sca2}
\end{equation}
and 
\begin{equation}
\langle R^{2}\rangle \sim t^{z},  \label{sca3}
\end{equation}
where $\eta $, $\delta $ and $z$ are the critical exponents associated to
the mean-number of predators, the survival probability, and the mean-square
distance of spreading, respectively.

\begin{figure}
\centering
\epsfig{file=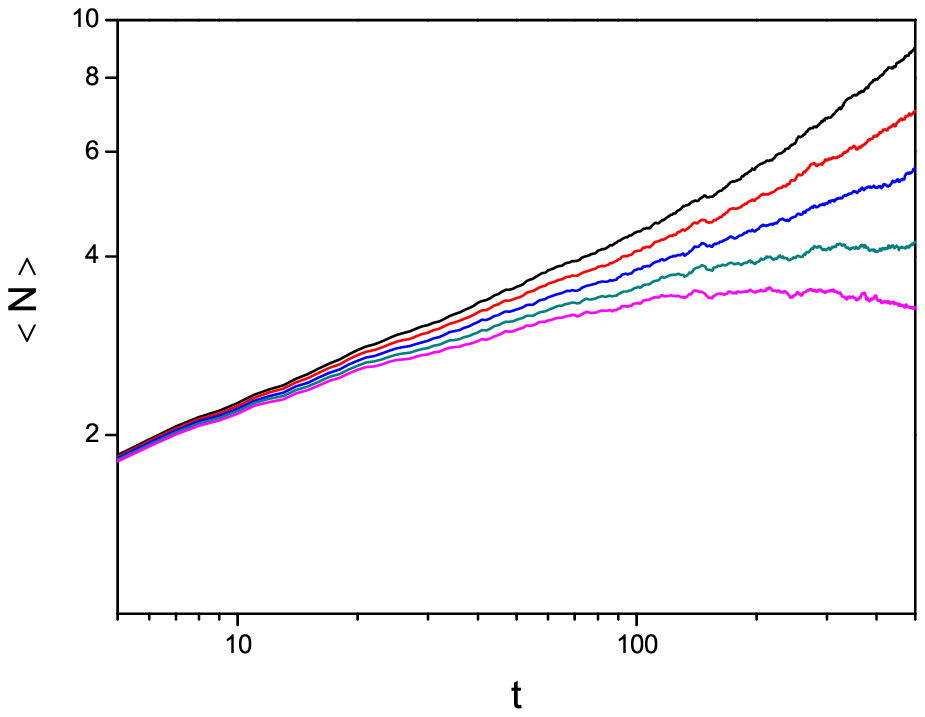,width=8.2cm}
\epsfig{file=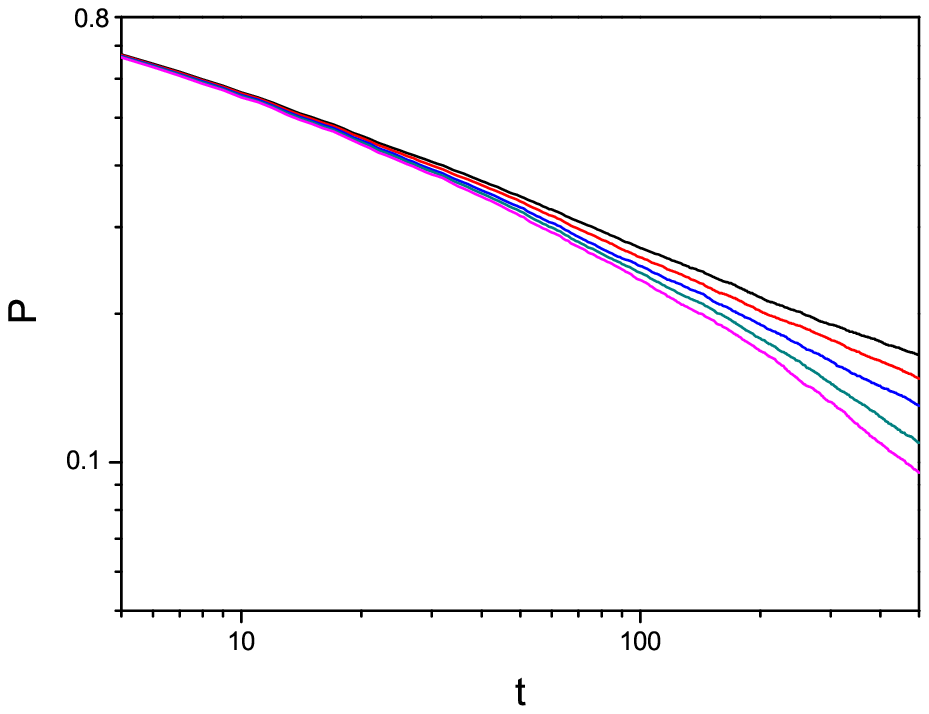,width=8.2cm}
\epsfig{file=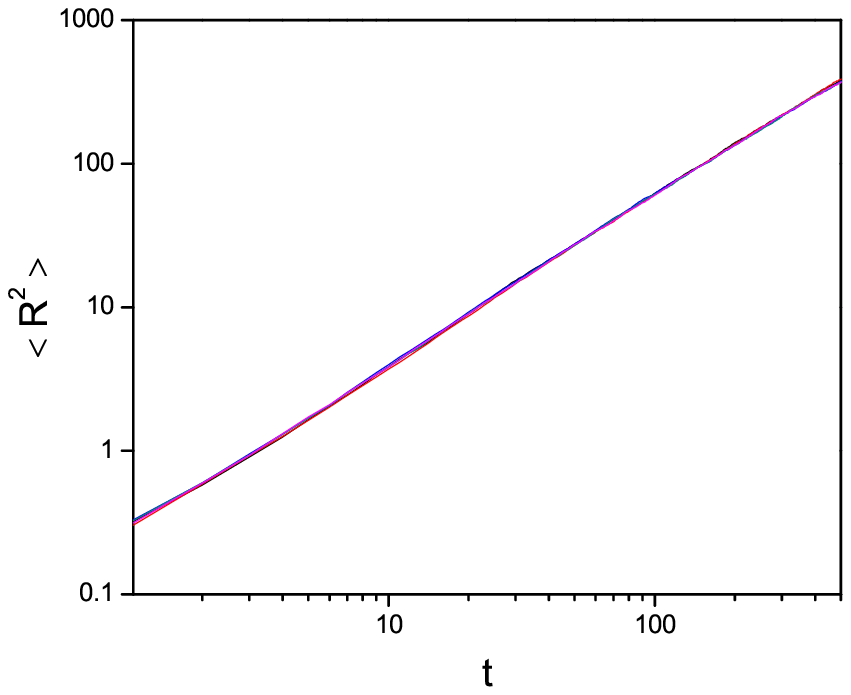,width=8.2cm}
\caption{Log-log plot of the mean-number of predators $\langle N\rangle $
(upper panel), the survival probability $P$ (middle panel) and the mean-
square distance of spreading of predators $\langle R^{2}\rangle $ (lower
panel), as a function of the time $t$, for $p=0$. Each figure shows the
behavior of these quantities for different values of $c$.
From top to bottom: $c=0.18875$, $c=0.18975$, $c=0.19075$, $c=0.19175$ 
and $c=0.19275$.}
\label{exponents}
\end{figure}

\begin{figure}
\centering\epsfig{file=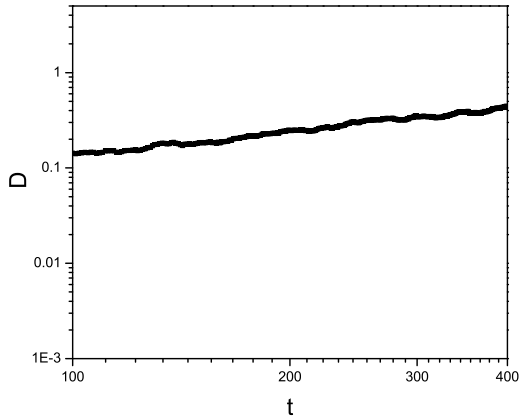,width=8.0cm}
\caption{Logarithmic derivative $D$ of $\langle N\rangle $ with
respect to $c$. For $p=0$ and considering values of $c$ slightly above 
and below the critical value.}
\label{derivada}
\end{figure}

As can be seen in Fig. \ref{exponents}, the log-log plot of the mean-number
of predators, as well as of the survival probability, show very well defined
critical and off-critical asymptotic behaviors. For a fixed value of $p,$
and values of $c$ below, but close to the critical value (supercritical
regime), these quantities present a positive curvature, which indicates the
presence of activity in the system; and for values of $c$ greater than the
critical value (subcritical regime) they present a negative curvature. At
criticality, $\langle N\rangle $ and $P$ exhibit a very clear asymptotic
power law behavior. 
Therefore, it is possible to obtain, with
precision, the phase
boundaries together with the critical exponents.
From Fig. \ref{exponents}
we get the following estimates for the critical exponents 
\[
\eta =0.230(2),\qquad \delta =0.451(3),\qquad z=1.134(4), 
\]
for $p=0$ (that is, for $a=b$). As reported in Table I, the same values for
the exponents, with the error bars, were obtained (using the same procedure)
for different values of $p$ along the critical line $c_{1}(p)$ (excluding
the critical point for which $a=0$). The errors of the exponents were
estimated by dividing $N_{S}=10000$ samples in bins of $2000$ runs, and
calculating the fluctuation of the averages obtained for each bin. These
values of the dynamic critical exponents are consistent with the ones of the
directed percolation class in $2+1$ dimensions: $\eta =0.2295(10)$, $\delta
=0.4505(10)$ and $z=1.1325(10)$ \cite{mun}.

\begin{table}
\caption{Phase boundaries and dynamic critical exponents for the
predator-prey cellular automaton. The values of the last column were
obtained using $\protect\beta =\protect\delta \protect\nu _{\parallel }$.}
{\footnotesize 
\begin{tabular}{lllllll}
\hline\hline
$p$ & $c$ & $\eta $ & $\delta $ & $z$ & $\nu _{\parallel }$ & $\beta $ \\ 
\hline
-0.495 & 0.00206(3) & 0.234(4) & 0.444(6) & 1.13(2) & 1.30(2) & 0.58(2) \\ 
-0.45 & 0.0228(3) & 0.227(4) & 0.448(4) & 1.13(1) & 1.31(2) & 0.59(1) \\ 
-0.40 & 0.04515(10) & 0.229(3) & 0.451(3) & 1.135(8) & 1.30(1) & 0.586(8) \\ 
-0.30 & 0.0877(2) & 0.226(3) & 0.452(3) & 1.134(7) & 1.29(1) & 0.583(8) \\ 
-0.20 & 0.12695(2) & 0.232(2) & 0.452(3) & 1.129(8) & 1.288(9) & 0.582(8) \\ 
-0.10 & 0.1617(2) & 0.234(3) & 0.447(3) & 1.135(6) & 1.295(10) & 0.579(8) \\ 
0.00 & 0.19075(10) & 0.230(2) & 0.451(3) & 1.134(4) & 1.292(8) & 0.583(7) \\ 
0.10 & 0.2127(2) & 0.230(2) & 0.449(3) & 1.131(7) & 1.287(8) & 0.578(7) \\ 
0.20 & 0.2243(2) & 0.228(2) & 0.450(3) & 1.128(8) & 1.291(9) & 0.581(8) \\ 
0.22 & 0.2253(2) & 0.231(2) & 0.447(3) & 1.131(7) & 1.29(1) & 0.58(1) \\ 
0.24 & 0.2262(2) & 0.230(3) & 0.445(5) & 1.136(9) & 1.30(1) & 0.58(1) \\ 
0.26 & 0.2263(2) & 0.226(4) & 0.443(6) & 1.129(8) & 1.29(1) & 0.57(1) \\ 
0.28 & 0.2262(2) & 0.232(4) & 0.444(6) & 1.135(10) & 1.29(1) & 0.57(1) \\ 
0.30 & 0.2257(1) & 0.231(4) & 0.445(5) & 1.13(1) & 1.295(10) & 0.58(1) \\ 
0.32 & 0.2250(2) & 0.233(4) & 0.448(5) & 1.14(1) & 1.30(1) & 0.58(1) \\ 
0.34 & 0.2239(2) & 0.234(5) & 0.449(6) & 1.13(1) & 1.29(1) & 0.58(1) \\ 
0.36 & 0.223(1) & 0.235(7) & 0.452(8) & 1.14(2) & 1.28(2) & 0.58(2) \\ 
0.38 & 0.223(2) & 0.24(1) & 0.44(1) & 1.14(2) & 1.29(2) & 0.57(2) \\ 
\hline\hline
$a$ & $c_{c}$ & $\eta $ & $\delta $ & $z$ & $\nu _{\parallel }$ & $\beta $
\\ \hline
0.00 & 0.220(3) & 0.587(2) & 0.096(4) & 1.767(5) & 1.51(1) & 0.145(7) \\ 
&  &  &  &  &  & 0.139(4)$^{\dag }$ \\ \hline\hline
\end{tabular}
$^{\dag }$See Sec. III D. }
\end{table}

\begin{figure}
\centering\
\epsfig{file=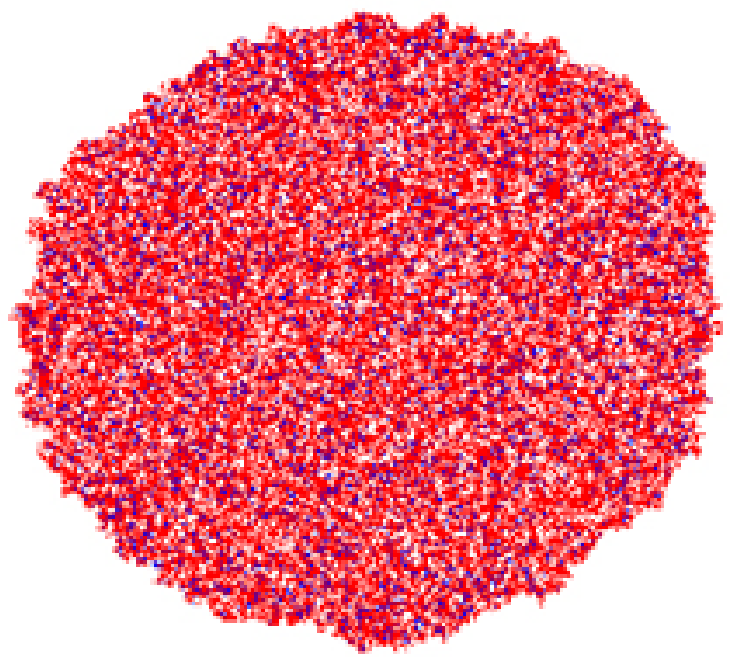,width=6.0cm}
\epsfig{file=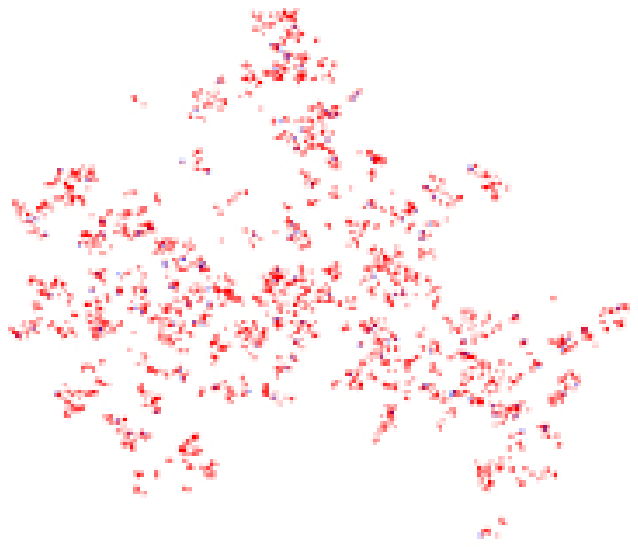,width=6.0cm}
\caption{Two snapshots of the predator-prey automaton, generated from a single
predator at the origin (center) of a lattice covered by
prey, for $p=-0.45$ and $c=0.01$ (upper panel, taken at $t=10000)$
and $c=0.0227$ (lower panel, taken at $t=65000$).
The white points represent sites occupied by prey, the red
points by predators and the blue points are empty sites.}
\label{pm45}
\end{figure}

\begin{figure}
\centering
\epsfig{file=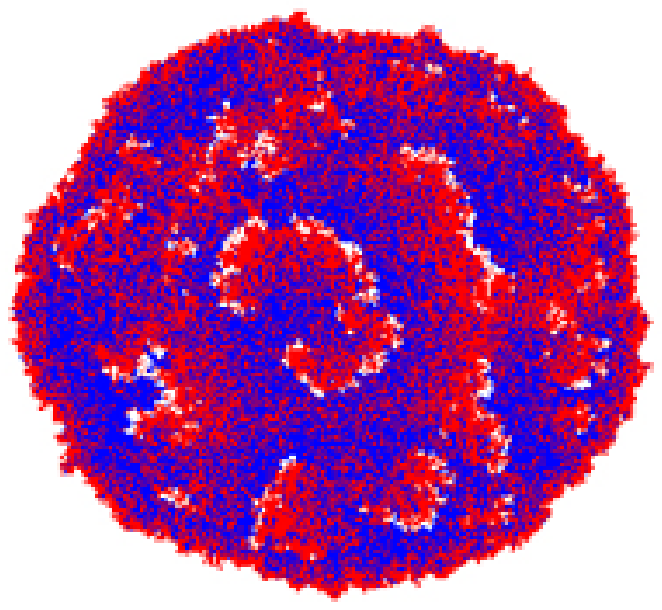,width=6.0cm}
\epsfig{file=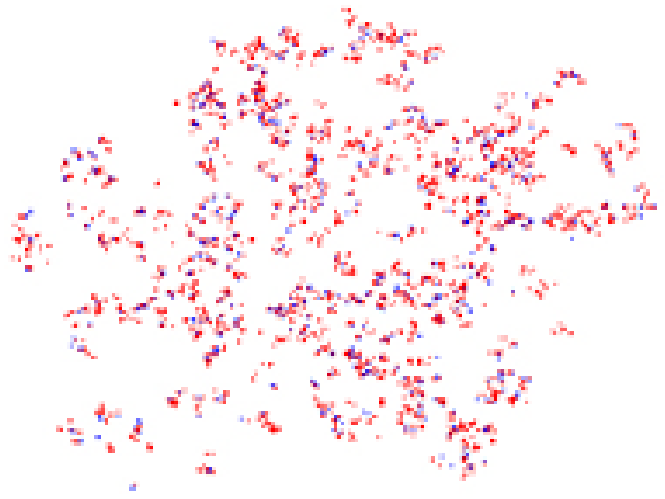,width=6.0cm}
\caption{The same of Fig. 4, for $p=-0.3$ and $c=0.0075$ (upper panel, taken
at $t=1750$), and $c=0.0867$ (lower panel, taken after $t=20000$). }
\label{pm30_act}
\end{figure}

\begin{figure}
\centering
\epsfig{file=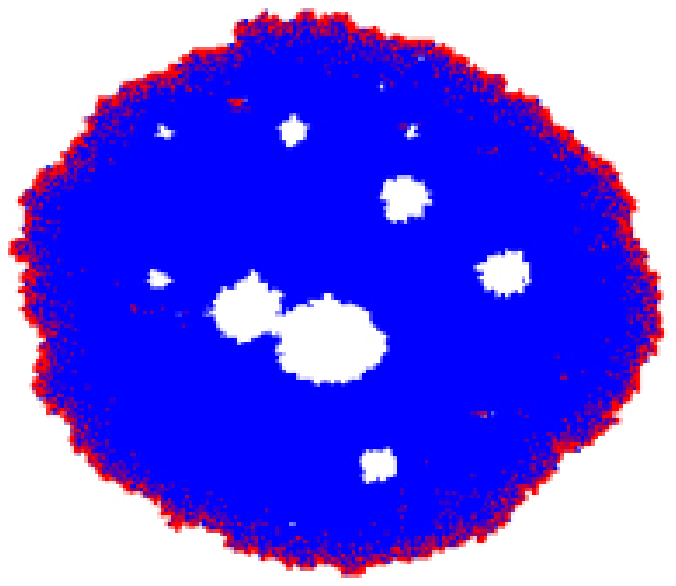,width=6.0cm}
\epsfig{file=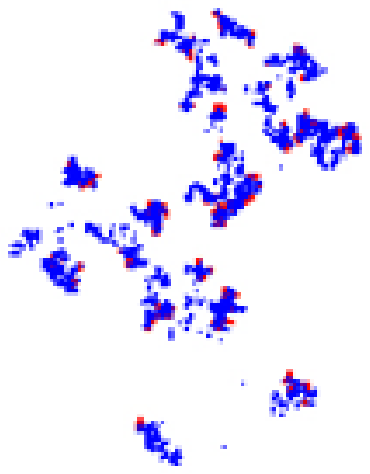,width=6.0cm}
\caption{The same of Fig. 4, for $p=0.3$ and $c=0.06$ (upper panel,
taken at $t=500)$, and $c=0.2257$ (lower panel, taken at $t=3000$).}
\label{p03_act}
\end{figure}

This dynamic critical behavior has been obtained for a diversity of models
with finite number of asymmetric absorbing states \cite{mun,hinrichsen},
such as the contact process, which can be viewed as a model for a simple
epidemic \cite{liggett}. The critical behavior of this model has been
extensively studied and it belongs to the directed percolation class 
\cite{jensen,hinrichsen,marro}.
Other examples are the Domany-Kinzel cellular
automaton \cite{domany} and the spreading of damage transitions
\cite{hinrichsen,grassberger4,etdm}.
Among the models with more then one
absorbing state, perhaps the most known is the Ziff-Gullari-Barshad model 
\cite{ziff} for the reaction of oxidation of carbon monoxide over a
catalytic surface; it also has dynamic critical exponents \cite{dickman1}
consistent with the ones associated to the directed percolation class.

The critical exponent associated to the time correlation
length $\nu_{\parallel }$ can be obtained from the
time-dependent simulations.
Here, the estimation was accomplished by studying the time behavior
of the derivative $D $ of the mean-value of
predators $\langle N\rangle $.
From the scaling laws it can be shown that \cite{grassberger3}, 
\begin{equation}
D=\frac{d\log \langle N\rangle }{d\log c}\simeq t^{\,1/\nu_{\parallel}},
\label{grass3}
\end{equation}
at the critical value of $c$. Data for different values of $c$, close to the
critical point, were taken in the same run and the derivatives calculated
considering finite differences; the behavior of $D$, for $p=0$, is shown in
Fig. \ref{derivada}. As reported in Table I, the value obtained for the
exponent $\nu _{\parallel }$, along the transition line, is in agreement
with the estimated value $\nu _{\parallel }=1.295(6)$ \cite{mun} for the
directed percolation in $2+1$ dimensions.

The static exponent $\beta $ associated to the order parameter
was calculated from the values for the
dynamic critical exponents $\delta $ and $\nu _{\parallel }$ and using
the scaling relation $\beta = \delta \nu _{\parallel}$
\cite{grassberger,hinrichsen}.
The resulting value for the exponent $\beta $, reported on Table I, is
consistent with directed percolation in $2+1$ dimensions \cite{mun}.

For sets of parameters such that $a$ approaches zero the critical
behavior of the model suggests a crossover to another universality class.
This feature will be explored in Sec. III.

\subsection{Growing clusters of predator-prey coexistence}

The pictures shown in Figs. \ref{pm45}, \ref{pm30_act}, and \ref{p03_act}
are related to configurations generated by simulations of the predator-prey
cellular automaton departing from the initial condition with one predator at
the origin and the lattice full of prey.
It can be seen in Fig. \ref{pm45} a
configuration in the supercritical regime (upper panel) and an almost
critical configuration (lower panel) for a set of parameters
which corresponds to $b<<a$ and $c<<a$
(left down corner of the triangular phase diagram of Fig. \ref{diag}).
In this case, predators stay in the lattice for a long period of
time, and when one of them dies it gives place to an empty site
that will be almost immediately occupied by a prey individual.
Consequently, just a few
number of empty sites are present in the steady state and
also in the growing clusters. This behavior suggests that
the present three state process, in the
limit $b\rightarrow 0$, $c\rightarrow 0$ with $b/c$ finite,
can be replaced by a two state process, similar to
the contact process, involving predators and prey,
as pointed out in reference \cite{tpb}.

The growing cluster of the upper panel of Fig. \ref{pm30_act} shows a
pattern formation related to an active region where prey and predators
coexist and can exhibit local time coupled oscillations. With a nonzero
probability, the dynamics will survive forever with the densities of each
species different from zero but less then $1$. This cluster, as well as the
supercritical growing clusters of Figs. \ref{pm45} and \ref{p03_act} (upper
panels), assumes after some time, which we call $\tau$, 
the asymptotic shape of a ball. 
The correlation length $\xi$ can be understood as the linear size of
the cluster at $t=\tau$ so that for $t>\tau$ the system is close
to the stationary active state.
Our results suggest that the radius $R$ of the ball
grows linearly with $t$ for $t>\tau$ 
in accordance with the shape theorem \cite{durrett0}.
At the critical point, $\tau$ and $\xi$ diverges.
Far from the critical point, the time $\tau$ the cluster takes to get
a defined shape can be very small.
For instance, for the set of parameters considered in the upper panel
of Fig. \ref{p03_act} the shape of a ball is already taken after just $200$
time steps. 

We notice that the supercritical cluster
of Fig. \ref{p03_act} corresponds to a configuration where
the populations of prey and
predators, under conditions of low densities, are grouped
into small clusters of a unique species that are isolated from each other, 
and coexist without extinction
in a highly heterogeneous space.
In this respect, this species coexistence can be compared to
the Huffaker's experiment, commented in Sec. I.
The coexistence for
the set of parameters of Fig. \ref{p03_act} is associated to local
self-sustained coupled time oscillations of prey and predators populations
and has been described in detail in reference \cite{tpb}.

\begin{figure}
\centering
\epsfig{file=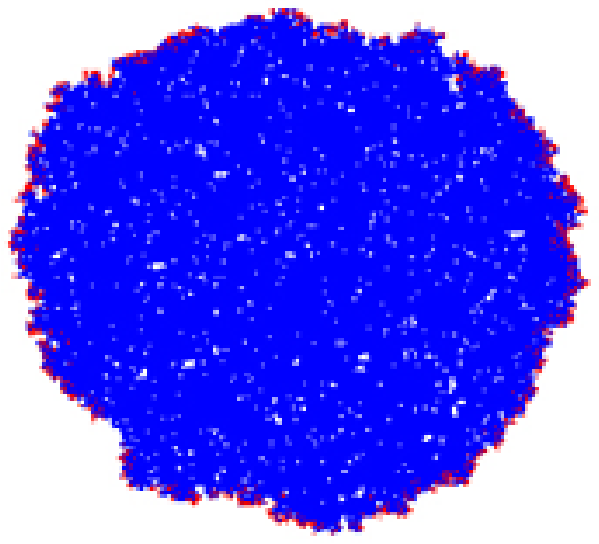,width=6.0cm}
\epsfig{file=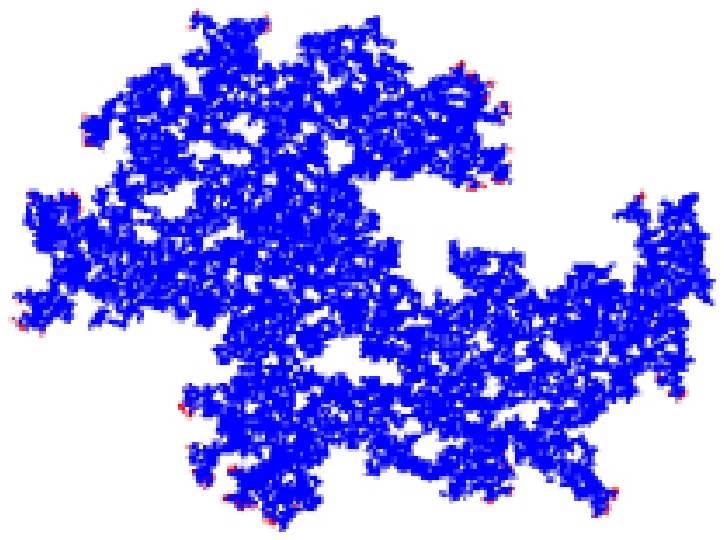,width=6.0cm}
\caption{Two snapshots of the epidemic with immunization ($a=0$) generated
from a single infected individual located at the origin (center) 
of a lattice covered by
susceptible individuals, for $c=0.15$ (upper panel, taken at $t=500$)
and for $c=0.22$
(lower panel, taken at $t=1000)$.
The white, red and blue points represent
sites occupied by susceptible, infected and an immune individuals,
respectively.}
\label{picture_a0}
\end{figure}

Typical clusters at criticality are shown in the lower panels
of Figs. \ref{pm45}, \ref{pm30_act} and \ref{p03_act}.
According to our results of Sec. II A,
these configurations must be related with transitions
belonging to the universality class of
directed percolation.
They do not present a ball shape, 
as the supercritical clusters described above,
but rather a shape of a fractal nature.
We note that the critical cluster of Fig. \ref{p03_act} (lower panel)
presents, in contrast with those of Figs. \ref{pm45} and \ref{pm30_act}
(lower panels), noticeable agglomerations of empty sites.
For values of $p$ corresponding to small values of $a$,
and, therefore, inside the crossover region, the critical
clusters present larger agglomerations of empty sites.
In the first time steps of the simulation, 
they resemble the critical growing cluster
for $a=0$ (lower panel of Fig. \ref{picture_a0}).
However, as long as the parameter $a$ is
different from zero, even if it is very small,
the critical cluster will present agglomerations of 
empty sites which can be always active, inasmuch as
prey can always proliferate in empty sites.
We remark that these critical clusters are still related
to transitions belonging to the directed percolation class.

A qualitative similar behavior of critical clusters was obtained by Dammer
and Hinrichsen \cite{hinrichsen1} for an epidemic spreading model which also
presents a crossover from the directed percolation critical
behavior to the dynamic percolation critical behavior.

\section{Spreading of an epidemic}

\subsection{The model}

For $a=0$\ the predator-prey probabilistic cellular automaton can be
interpreted as a model for the propagation of an epidemic with
immunization.
It mimics the spreading of an epidemic in a population 
composed by susceptible individuals ($X$) that become
infected ($Y$) by contact with infected
individuals; once infected the individuals
can recover, and become immune ($Z $) spontaneously.
The process of infection can be represented by the
reaction: 
\begin{equation}
X+Y\rightarrow 2Y,  \label{epid1}
\end{equation}
and the recovery process, which includes immunization, by the reaction: 
\begin{equation}
Y\rightarrow Z.  \label{epid2}
\end{equation}
These are the basic processes which are taken into account in the modeling
of the spreading of an epidemics with
immunization \cite{grassberger1,grassberger2,hastings,bocara,hinrichsen1}.
The present model
is defined on a regular square lattice where each site can be in the
following states: occupied by a susceptible, an infected or an immune
individual. It comprehends the following stochastic rules:

(A) The infection can occurs when a susceptible individual, which occupy a
given site, has at least one site occupied by 
an infected individual in its neighborhood, reaction (\ref{epid1}).
This process occurs with probability $b/4$
times the number of infected individuals in the neighboring sites.

(B) The recovering process can occurs spontaneously with probability $c$
when a site is occupied by an infected individual, reaction (\ref{epid2}).
The condition $b+c=1$ is obeyed, with $b$ being the infection probability
and $c$ the recovery probability.
This model can exhibits infinitely many
absorbing states and presents a continuous transition belonging to the
dynamic percolation universality class, as we show in Sec. III C.

\begin{figure}
\centering
\epsfig{file=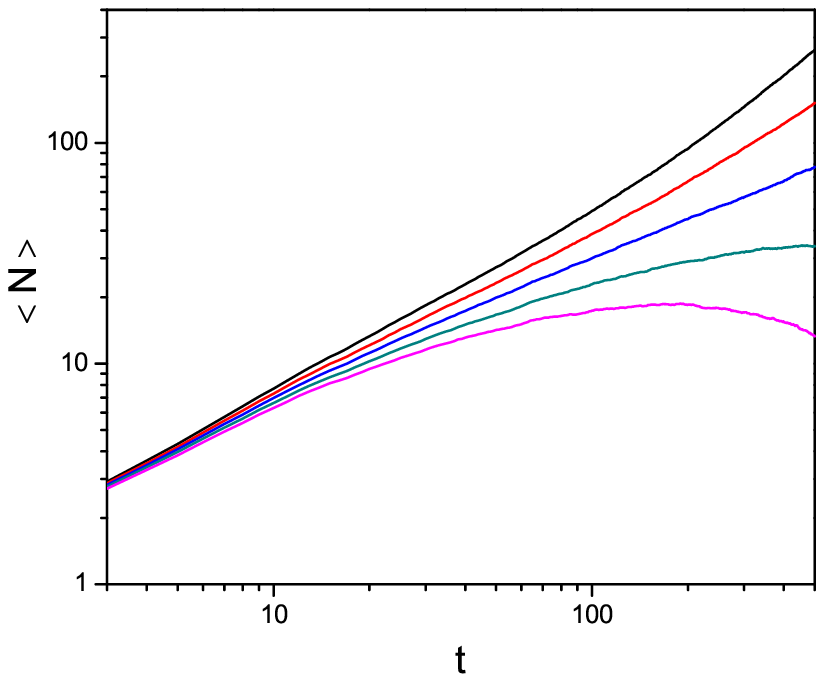,width=8.0cm}
\epsfig{file=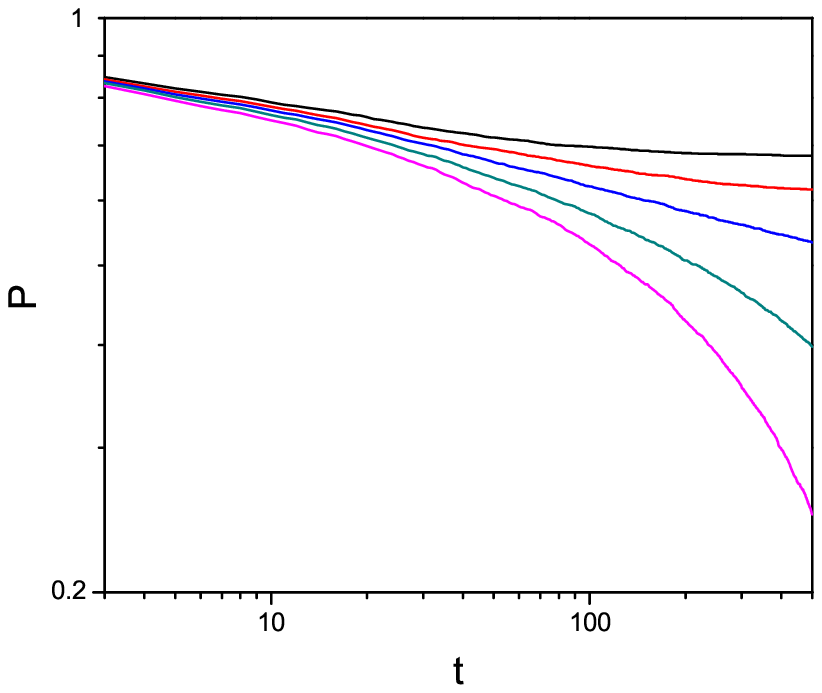,width=8.0cm}
\epsfig{file=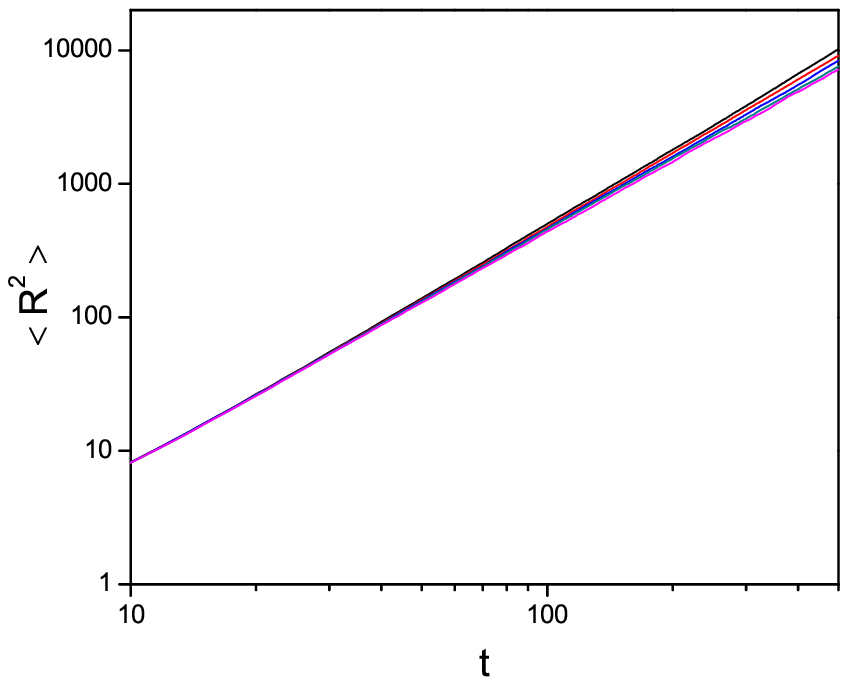,width=8.0cm}
\caption{Log-log plot of the mean-number of infected (upper panel), the
survival probability (middle panel) and the mean-square distance of
spreading of infected (lower panel), for $a=0$.
Each figure shows the
behavior of the quantities for different values of $c$.
From
top to bottom: $c=0.210$, $0.215$, $0.220$, $0.225$ and $0.230$.}
\label{exponents_a0}
\end{figure}

\begin{figure}
\centering
\epsfig{file=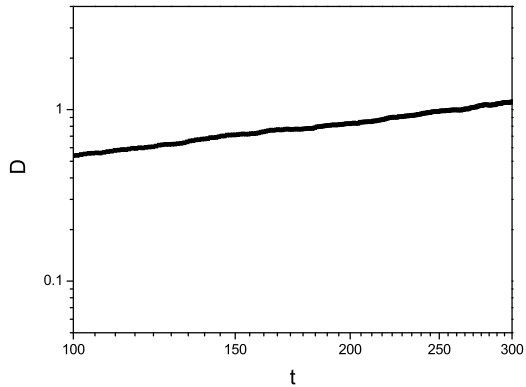,width=8.0cm}
\caption{Logarithmic derivative $D$ of $\langle N\rangle $ with
respect to $c$.
For $a=0$ and considering values of $c$ close to the critical value.}
\label{derivada_1}
\end{figure}

\subsection{Patterns of spreading}

To characterize the threshold of the epidemic spreading we perform
time-dependent simulations. Initially all sites of the lattice are
occupied by susceptible individuals except one, in the origin (center) of the
lattice, which is occupied by an infected individual. The stochastic 
dynamics follows the rules (A) and (B) defined in Sec. III A 
with a synchronous updating. The infection starts to be transmitted 
at time $t=0$. The problem is to know how an epidemic propagates 
(or not) in the population.
For high values of the probability $b$ of infection compared to the values
of the recovering probability $c$, the epidemic spreads leaving a cluster of
inactive sites composed by immune individuals and some groups
of individuals which remain susceptible forever.
A typical picture of the lattice in this
situation in shown in the upper panel of Fig. \ref{picture_a0}.
The cluster grows with a front of infected individuals which stay
in the border.
After a finite time $\tau$, the cluster assumes a limiting shape of a
ball which spreads to infinity with a nonzero probability.
We observe that different initial seeds lead to
different configurations of the spreading of the epidemic,
giving rise to an infinitely many absorbing states.
As $b$ is decreased (and $c$ is increased)
the threshold for the spreading of the epidemics is reached.
Above the threshold, the epidemic will cease in a finite time
leaving a cluster with just a few number of immune individuals,  
the rest of the lattice being covered by susceptible individuals.
A picture of the lattice near the threshold of
epidemic spreading is shown in the lower panel
of Fig. \ref{picture_a0}.
This almost critical cluster presents an irregular shape 
of fractal nature.

Similar critical and supercritical clusters have been
obtained from time-dependent simulations
for other spatial-structured stochastic models for
an epidemic with immunization \cite{durrett0,hinrichsen,hinrichsen1}.

\begin{figure}
\centering
\epsfig{file=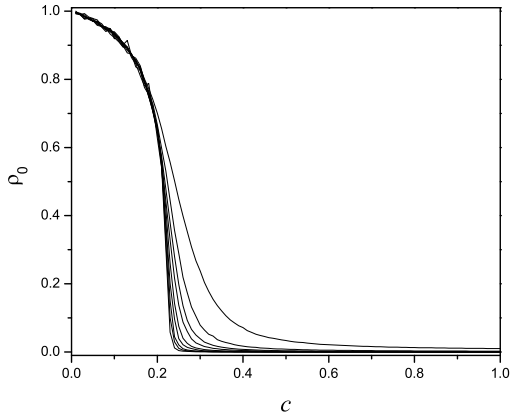,width=8.0cm}
\caption{Density of immune individuals $\protect\rho _{0}$ as
a function of $c$, for $a=0.0$.
From top to bottom (at right): $L=10, 20, 30,40,$
$60$, $80$, $120$, $160$, $240$. 
The critical value of $c$ for the
infinite system is $c_{c}=0.220(3)$ (see Table I).}
\label{vazios_c}
\end{figure}

\begin{figure}
\centering
\epsfig{file=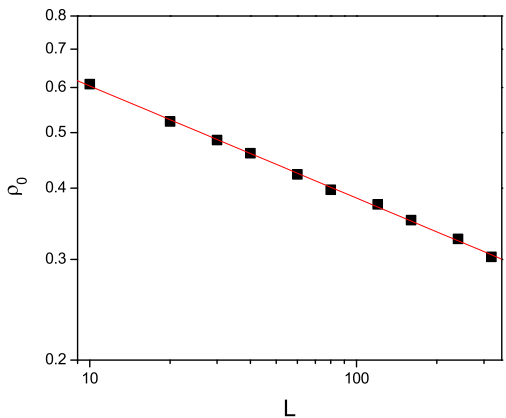,width=8.0cm}
\caption{Log-log plot of $\rho _{0}$,
calculated at the critical value $c_{c}=0.220$,
for $a=0.0$, as a function of $L$.
Simulations were performed on square lattices
of size ranging from $L$ $=10$
until $320$, and considering periodic boundary conditions.}
\label{scaling}
\end{figure}

\subsection{Dynamic critical exponents}

In Fig. \ref{exponents_a0} it is shown the time behavior of the mean-number
of infected individuals $\langle N\rangle $, the survival probability $P$,
and the mean-square distance of spreading from the origin $\langle
R^{2}\rangle $ obtained from time-dependent simulations performed near the
critical point. Since the log-log plots of of these quantities show very
clear power-law behavior, associated to critical behavior, then the
dynamic critical exponents $\eta $, $\delta $ and $z$, defined
in Eqs. (\ref{sca1}), (\ref{sca2}) and (\ref{sca3}), 
can be obtained with precision. The estimated critical exponents are
reported in Table I. These exponents are consistent with the critical
exponents of the dynamic percolation universality class \cite{mun}.
This is the expected critical universal behavior for a
general epidemic with
immunization \cite{grassberger1,grassberger2}.

The critical exponent $\nu _{\parallel }$ was obtained by analyzing the
behavior of the derivative $D$, defined in Eq. (\ref{grass3}), at the
critical point. From Fig. \ref{derivada_1}
we obtain the estimation $\nu_{\parallel }=1.51(1)$ which is consistent
with the value  $\nu _{\parallel }=1.506(1)$ \cite{mun}
of the dynamic percolation in two dimensions.
The static critical exponent $\beta $ associated
to the order parameter can be
evaluated from the estimates for the dynamic
critical exponents using the scaling 
relation $\beta =\delta \nu _{\parallel }$.
The resulting value, which is
given in Table I, is in agreement with the one for the dynamic percolation 
\cite{mun}.
We also have estimated this exponent by performing steady state
simulations in finite systems, as described in the next section.

\subsection{Critical exponent $\beta$}

To obtain another estimation of the critical
exponent $\beta $ associated to the order parameter
we have considered square lattices with linear size $L$ and
periodic boundary conditions.
For each value of $L$ we have performed
several independent simulations runs all 
starting from a lattice covered
with susceptible individuals with the exception
of one infected at the center of the lattice.
The system evolves in time according to the synchronous
stochastic dynamics defined in Sec. III A. In principle,
any one of the infinitely many absorbing states can be reached; 
and, consequently, the number of
immune individuals at the steady state varies
from sample to sample. A simulation is finished when the
system enters in an absorbing state and then the number of immune
individuals is calculated. The mean value of this quantity, divided by the
total number of lattice sites $L^{2}$, is the density $\rho _{0}$ of immune
individuals at the steady state (for a given lattice size).

The behavior of $\rho _{0}$ as a function of $c$,
for different values of $L$ is shown in
Fig. \ref{vazios_c}. In the subcritical regime ($c$ above its critical
value $c_c$) $\rho _{0}$ decreases as $L$ is increased and
in the limit $L\rightarrow \infty $ it vanishes.
In the supercritical regime $\rho _{0}$ is almost independent of $L$.
We have assumed that $\rho _{0}$ is an appropriate order parameter
for this transition.
Assuming the finite size scaling hypothesis \cite{marro}
we expect that $\rho _{0}(\Delta ,L)$, 
where $\Delta=c-c_c$, calculated for each
value of $L$, scales at the critical point as,
\begin{equation}
\rho _{0}(0,L)\sim L^{-\beta /\nu _{\perp }},
\end{equation}
where $\nu_{\perp}$ is the critical exponent related to the spatial
correlation length.

To obtain the ratio $\beta /\nu _{\perp }$ we plot $\log \rho _{0}$,
calculated at $c_{c}$, versus $\log L$, as shown in Fig. \ref{scaling}.
We obtain $\beta /\nu _{\perp }=0.185(5)$.
Using the result $\nu _{\perp}=4/3$ \cite{mun} we
find the value $\beta =0.139(4)$ which is in agreement
with the one for the isotropic percolation, namely $\beta =5/36$ \cite{mun}.
And, also in agreement with the value estimated in Sec. III C.

\section{Summary}

We have studied a spatial-structured model in which prey and predators
individuals reside on the sites of a lattice and are described by discrete
dynamic variables. The system evolves in time according to a probabilistic
cellular automaton which takes into account the Lotka-Volterra interactions
by the use of Markovian local rules. The model exhibits
an active phase where prey and predator coexist without
extinction, an absorbing prey phase, and, when the
birth probability of prey vanishes, a phase with infinitely many
absorbing states composed by empty (immune) sites and prey (susceptible).
In the last case, it is more appropriate to refer to the predator-prey cellular
automaton as a model for the spreading of an epidemic with immunization in a
population composed by susceptible, infected and immune individuals.
The automaton was studied by time-dependent simulations which allow us to
determine with precision the phase boundaries (thresholds) and the dynamic
critical exponents. We have localized a transition line which crosses the
entire phase diagram. For all sets of parameters such that the prey birth
probability is different from zero, this line separates the active phase,
where prey and predators coexist, and the prey absorbing phase. We have
shown that this transition belongs to the directed percolation
universality class.
We also have shown that when the prey birth probability equals
zero, there is a crossover to the universality class of the dynamic
percolation.

Patterns of growing clusters with coexistence of prey, predators and empty
sites (or, susceptible, infected and immune individuals), generated by the
time dependent simulations, in the critical and in the supercritical
regimes, were shown and discussed. The present study provides a detailed
description of the thresholds of stable coexistence of the two species (or,
the spreading of an epidemics) in the context of this predator-prey cellular
automaton.

\section*{Acknowledgements}

The authors have been supported by the Brazilian agency CNPq.


\end{document}